\documentclass[journal]{vgtc}                     

\onlineid{1394}

\vgtccategory{Research}

\vgtcpapertype{application/design study}

\title{
Discursive Patinas: Anchoring Discussions in Data Visualizations
}

\author{%
  \authororcid{Tobias Kauer}{0000-0003-0746-904X},
  \authororcid{Derya Akbaba}{0000-0001-9419-3402},
  \authororcid{Marian Dörk}{0000-0002-3469-7841}, and
  \authororcid{Benjamin Bach}{0000-0002-9201-7744}
}

\authorfooter{
  \item Tobias Kauer (t.kauer@ed.ac.uk) is with University of Edinburgh (UK) and the Potsdam University of Applied Sciences (DE)
  \item Derya Akbaba is with Linköping University (SE)
  \item Marian Dörk is with Potsdam University of Applied Sciences (DE)
  \item Benjamin Bach is with Inria (FR) and the University of Edinburgh (UK)

}

\abstract{This paper presents discursive patinas, a technique to visualize discussions onto data visualizations, inspired by how people leave traces in the physical world. While data visualizations are widely discussed in online communities and social media, comments tend to be displayed separately from the visualization and we lack ways to relate these discussions back to the content of the visualization, e.g., to situate comments, explain visual patterns, or question assumptions. In our visualization annotation interface, users can designate areas within the visualization. Discursive patinas are made of overlaid visual marks (\emph{anchors}), attached to textual comments with category labels, likes, and replies. By coloring and styling the anchors, a meta visualization emerges, showing what and where people comment and annotate the visualization. These \emph{patinas} show regions of heavy discussions, recent commenting activity, and the distribution of questions, suggestions, or personal stories. We ran workshops with 90 students, domain experts, and visualization researchers to study how people use anchors to discuss visualizations and how patinas influence people's understanding of the discussion. Our results show that discursive patinas improve the ability to navigate discussions and guide people to comments that help understand, contextualize, or scrutinize the visualization. We discuss the potential of anchors and patinas to support discursive engagements, including critical readings of visualizations, design feedback, and feminist approaches to data visualization.}


\teaser{

  \includegraphics[width=\textwidth]{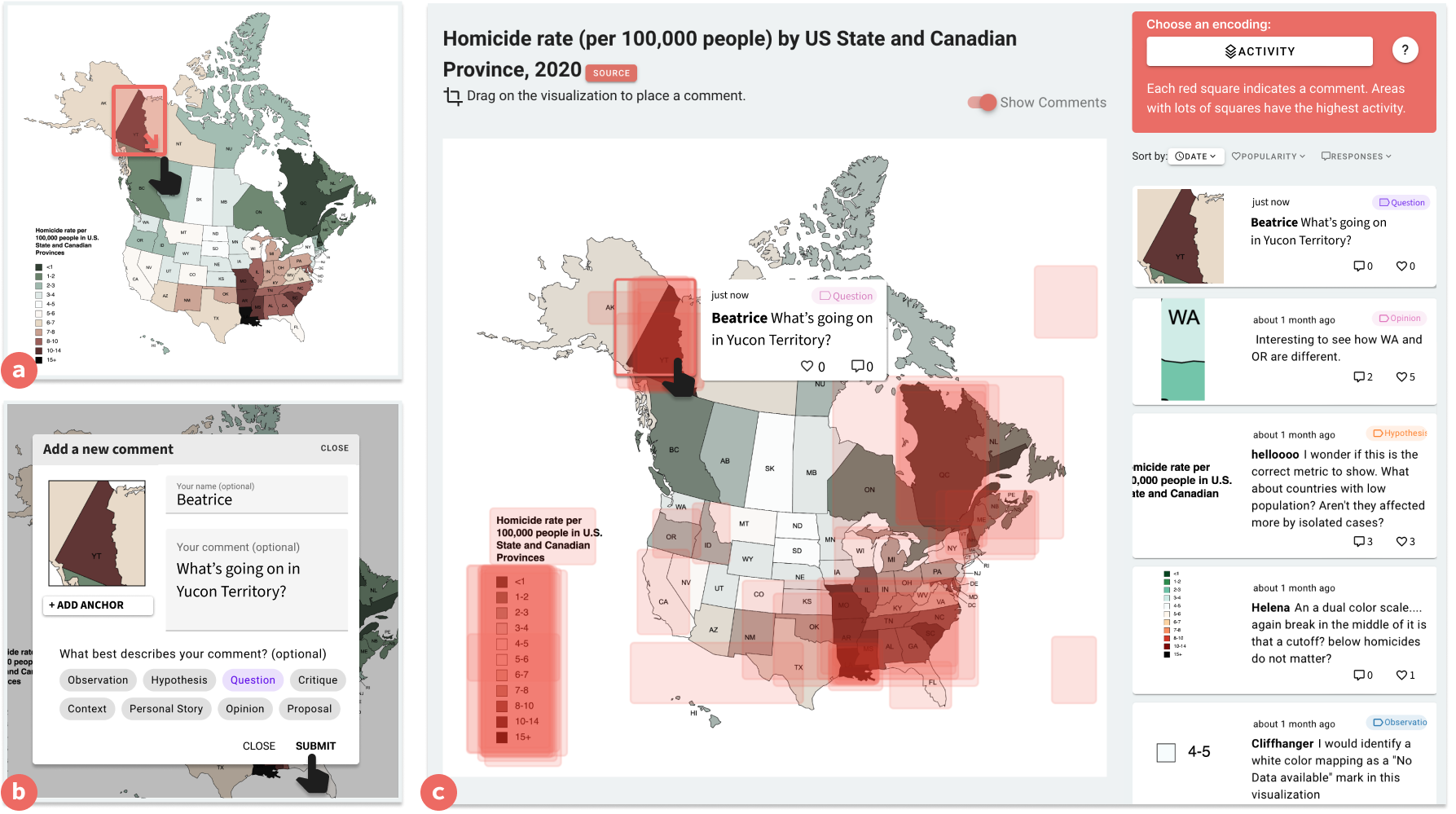}
  \caption{\label{fig:authoring}\revised{Our platform supporting discussions about visualizations through anchored comments (a, b) and discursive patinas (c). Drawing on the visualization creates an anchor (a) and a textual comment (b). The full  interface shows all anchored comments as an emerging discursive patina on the visualization. All comments are also shown in a separate browsable list (c). Together, anchored comments and discursive patinas visualize the level of engagement of specific areas in a visualization that are being discussed across people.}}
}

\graphicspath{{figs/}{figures/}{pictures/}{images/}{./}} 

\usepackage{tabu}                      
\usepackage{booktabs}                  
\usepackage{lipsum}                    
\usepackage{mwe}                       

\usepackage{mathptmx}                  

\usepackage{multirow}
\usepackage{tabularx}
\usepackage{color,soul}
\usepackage{dblfloatfix}

\newcommand{\revised}[1]{#1}

\definecolor{comment}{RGB}{253,234,233}
\definecolor{survey}{RGB}{255,245,228}
\definecolor{statement}{RGB}{237,242,255}

\newcommand{\commentsymbol}{\raisebox{-.2ex}{\includegraphics[height=1.3\fontcharht\font`\B]{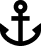}}}
\newcommand{\statementsymbol}{\raisebox{-.3ex}{\includegraphics[height=1.3\fontcharht\font`\B]{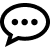}}}
\newcommand{\surveysymbol}{\raisebox{-.2ex}{\includegraphics[height=1.3\fontcharht\font`\B]{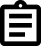}}}

\DeclareRobustCommand{\comment}[2]{\textit{``#1''}~(\commentsymbol{}~\texttt{#2})}
\DeclareRobustCommand{\statement}[2]{\textit{``#1''}~(\statementsymbol{}~\texttt{#2})}
\DeclareRobustCommand{\survey}[1]{\textit{``#1''}~(\surveysymbol{}~\texttt{CE})}

\begin{document}
\firstsection{Introduction}
\maketitle
\newcommand{\platform}{\textit{Viscussion}}

From politics to popular culture, data visualizations have become a common method to communicate contemporary topics to broad audiences. Visualizations can inform a wider public on social media, news websites and community platforms, and spark discussions on a wide range of issues. It is often the discussions unfolding around visualizations that support sensemaking~\cite{heer2007voyagers, viegas2007manyeyes, willett2011commentspace}, help scrutinize data~\cite{hullman2015content}, ask for context or critique~\cite{walny2020pixelclipper,lin2022data}, or invite people to share personal stories and experiences~\cite{kauer2021public,peck2019data}. Several research prototypes and platforms have been developed to facilitate discussions with data visualization, such as Sense.us~\cite{heer2007voyagers}, Many Eyes~\cite{viegas2007manyeyes}, and CommentSpace~\cite{willett2011commentspace}. These environments offer a range of features to annotate visualizations, leave textual messages, and reply to comments. 

We see two major approaches to visualization discussion platforms. The first one focuses on textual comments that are displayed separate from the visualization, i.e., comments exist as individual text snippets besides or underneath the visualization with no explicit relationship to any part of the visualization~\cite{kauer2021public,hullman2015content}. Such interfaces are simple and scalable as long as the comments refer to the visualization as a whole---however, they do not support detailed comments about specific visual elements or data patterns in the visualization and do not relate comments back to the visualization~\cite{he2023enthusiastic,hullman2015content,mcinnis2020rare}. For example, a comment might critique the scale on the legend or highlight a specific visual pattern in the visualization. To mitigate this problem, the second approach allows for free-form annotations where users can highlight and otherwise express themselves directly \textit{on} the visualization~\cite{heer2007voyagers,lin2022data,romat2019activeink}. While allowing for more direct comments, such free-form annotations can quickly clutter the visualization, especially in scenarios with hundreds of commenters, as is common on online platforms. Moreover, existing interfaces hardly convey the complex, diverse, and often highly dynamic nature of discussions. It is difficult to get an overview about what people are discussing, what they are saying, and which parts of a visualization spark most conversation.

We introduce an annotation interface designed to scale the number of comments about a visualization while closely relating the comments to the aspects in the visualization they refer to.
The web-based interface---\platform---
supports \textit{anchored comments} that can be placed onto the visualization using semi-transparent rectangles as visual marks.
In contrast to traditional map markers (e.g., pins), these rectangles designate relevant areas in a visualization and can be styled to encode information about a comment. This combination of visual encoding and scope of comments summarize aspects of the discussion within the spatial structure of the visualization.
\revised{Similar to existing work on crowdsourced design critique~\cite{Luther2015}, our} 
intent is to foster constructive and critical discourse that is literally grounded in the data visualization.
As the number of anchors and comments grows, meta-visualizations emerge that we call \textit{discursive patinas} (see \autoref{fig:authoring}), whose designs are inspired by visualizing interaction histories~\cite{matejka2013patina} as well as traces human interactions can leave on objects in the physical world~\cite{hill1992edit, wexelblat1999footprints, dieberger2000social, brusilovsky2008social}. Discursive patinas are designed \textit{``to guide [...] actions, to make choices, and to find things of importance or interest''} \cite{wexelblat1999footprints} within a visualization.

\revised{Across three complementary studies (Sec.~\ref{sec:evaluation}) involving a total of 90 participants, we use \platform{} in different contexts to} investigate the potential of anchored comments and discursive patinas. 
We study how discussions on data visualizations can serve not only as means for understanding data but also as a communicative space for critical reflection about biases in the data and assumptions in the visualization design~\cite{dork2012critical,dignazio2023datafeminism}.
The results from these studies indicate that anchored comments can reduce barriers to understanding a visualization, provide context, and introduce alternative interpretations. Compared to the baseline interface, discursive patinas improved participants' navigation of the discussion by highlighting different areas of interest within the visualization. Overall, anchored comments and discursive patinas encouraged people to more carefully examine specific parts of a visualization, compared to when they only had the ability to create unanchored comments. Based on these results we discuss how discursive patinas can support \textit{close reading of visualizations} while at the same time allowing for \textit{distant reading of the associated discussions}. We conclude by discussing future scenarios for discursive annotation interfaces that allow for critical readings of visualizations, provide design feedback, and support feminist approaches to data visualization.

In summary, we contribute a technique for integrating and aggregating discussions on data visualizations and report from a three-step evaluation on the impact of anchored comments and discursive patina on critical engagement with data visualization.

\section{Background}\label{sec:RW}
Our research broadly relates to participatory mapping and visualization, and specifically builds on prior work on visualization annotation and interaction histories.

\subsection{Community mapping and discussing visualizations}
Maps have been used extensively for participatory data collection and mapping activities~\cite{orangotango2018this}, owing their potential to set up spaces for people to meet, exchange, and share claims about the world~\cite{loukissas2021open}. In such projects, users write and collect often personal stories, e.g., about housing evictions or queer moments while locating them on a web-based map~\cite{maharawal2018anti,brown2008queering}. 
\revised{The annotations represent \textit{``stories for someone''}, i.e., stories not addressed to anyone specifically, but written to create an archive of experiences~\cite{kirby2021queering}. }

Data visualizations, like geographic maps, can arguably play a similar role in such endeavours, e.g., by providing evidence about a topic of discussion, or by offering a canvas to solicit and locate stories, opinions, expressions, thoughts, questions, and more~\cite{kauer2021public}. Some studies reported that people seek entertainment to satisfy their curiosity, or to avoid boredom when interacting with casual visualizations \cite{sprague2012exploring, pousman2007casual}. Other studies point to different aspects of identity and lived experiences affecting engagement with visualizations on a deeply personal level, pushing the beholder towards visualizations that reflect their own experiences \cite{peck2019data}. Additionally we've begun to study how prior beliefs impact how people see correlations in the data~\cite{xiong2022seeing} and whether they trust political analyses~\cite{yang2023swaying}. 
Most recently, Kauer et al.~\cite{kauer2021public} created a taxonomy of user reactions from Reddit comments, citing reactions such as observations, proposals, personal stories, or additional information.
Our goal is to make the diversity of comments accessible to readers of visualizations and those who engage in the discussions. 

Further studies investigating visualization comments on online news platforms, however, found that the majority of comments ignore the data or visualization itself. Instead, comments focus on the context and framing of the story~\cite{hullman2015content} while only a fraction of comments refer to the data visualizations in the news articles~\cite{mcinnis2020rare}. 
This connects with the observation that discussions \textit{``generally integrate poorly with visualization and analysis tools''}~\cite{he2023enthusiastic} since comments are displayed in separate sections~\cite{kauer2021public, hullman2015content, mcinnis2020rare} or use \textit{``screenshots and external links rather than more malleable representations''}~\cite{he2023enthusiastic}. In this paper, we offer a mechanism to embed the discussion of---and on---the visualization.

\subsection{Visualization annotation interfaces}
While placing comments outside of visualizations remains a simple interface design solution, several research projects have explored how to integrate comments more closely with visualizations. 
Sense.us~\cite{heer2007voyagers} invites collaborators to comment on data visualizations asynchronously, with comments displayed separately from the visualization. Comments can be linked to freehand drawings, shapes or text that are superimposed on the visualization. Expanding on this idea, CommentSpace~\cite{willett2011commentspace} provides categories for comments (e.g., hypothesis, evidence-for). 
With PixelClipper, readers can create custom cutouts of a visualization based on a grid and write comments about these portions of the visualization~\cite{walny2020pixelclipper}. Cutouts are then shown alongside textual comments to provide a reference to the visualization content\revised{. Multitudes of cutouts are not shown overlaid on the visualization, making it difficult to find areas of high-comment traffic}. 

More direct integration of annotations is supported by free-form comments. For example, 
ActiveInk~\cite{romat2019activeink} investigates techniques to draw and write directly on visualizations and to highlight, hide or manipulate visual marks with digital pens. Studies found analysts highly appreciate such direct annotations~~\cite{kim2019inking}. 
While very expressive, these techniques focus on personal annotations, rather than collaborations with potentially many users. 
Annotations can help identify discrepancies in data or add experts' personal knowledge to charts to indicate potential flaws in the represented data. 
Highlighting relevant data points or areas in a chart can help the beholder navigate and better understand the information~\cite{romat2019activeink, kim2019inking}. 
For example, markers on maps can help externalize and discuss implicit errors in visual analytics systems~\cite{mccurdy2018}. Furthermore, free-form drawing could aid in linking comments to the specific parts of a visualization they refer to~\cite{lin2022data}. 

The discussed approaches introduce various strategies to embed annotations within the visualization. Our contribution is characterised by an assessment of prior annotation interfaces \revised{focusing on two axes of functionality:}
scale and scope. 
The concept of \textit{scale} reflects the interfaces' ability to accommodate a growing volume of comments without leading to visual clutter or compromising the readability of individual comments. Conversely, by \textit{scope}, we refer to the mechanism for identifying portions of a visualization that are being commented on.
While early work offers no visual integration at all~\cite{viegas2007manyeyes}, later tools use sophisticated mechanisms to place anchors, e.g., by allowing users to create text~\cite{heer2007voyagers, romat2019activeink}, freehand drawings \cite{churchill2000anchored,heer2007voyagers, romat2019activeink, lin2022data}, or by manipulating existing marks on their original location~\cite{lin2022data, romat2019activeink, willett2011commentspace}.
With these tools, anchors can effectively encircle, point to, highlight, or cut out portions of visualizations they reference. The high flexibility in defining the scope comes at the expense of their capacity to scale with increasing numbers of contributions. Overlapping freehand drawings render each other illegible; for a similar reason clippings~\cite{walny2020pixelclipper} or manipulations~\cite{romat2019activeink, lin2022data} from multiple viewers are not presented in one view. While this is not an issue for individuals or small groups, in public forums it quickly becomes an issue of scale when idiosyncratic scope definitions can impede the overall legibility.

In contrast to free-form annotations, interfaces using one-dimensional markers \revised{on maps or images}~\cite{brown2008queering, mccurdy2018, Xu2014} are highly scalable, at the cost of custom scope. The uniformity and small size of markers help readers to quickly gauge where many comments are placed. Their identical size, however, can make it hard to understand the scope of what a comment refers to. 
With our research, we are particularly interested in resolving the tension between scale and scope of visualization annotations.

\subsection{Visualizing user behavior}
Prior work on visualizing user behavior within interfaces offers cues for how to address the question of scale and scope. 
Akin to \textit{footsteps} forming a path, there has been a long-standing interest in logging and displaying user interactions within web applications to generate social signals~\cite{hill1992edit, wexelblat1999footprints, dieberger2000social, brusilovsky2008social} that provide future users with information about past users' behavior. 
Interaction histories have been visualized to provide \textit{proximal cues} that help people find useful information~\cite{pirolli1999information}. Scented widgets~\cite{willett2007scented}, for example, add visual indicators, such as bar charts, to interface elements to enhance social navigation by recording and representing visitation patterns, indicating popular or neglected items within an information system.
The idea of embedding visual cues to convey prior use can be extended to any kind of application without requiring any instrumentation, by superimposing patina heatmaps that aggregate application usage data~\cite{matejka2013patina}.
HindSight~\cite{feng2016hindsight} translates the footsteps analogy to visualization and changes the color of elements in a visualization after the user has visited it, thus aiding the personal analysis process.

Visualizing prior interaction data can change how users explore interfaces. Personal interaction histories encourage users to engage more deeply with presented data, as they increase the volume of data they consider~\cite{feng2016hindsight}. Visually supported social navigation aids users in exploring unfamiliar datasets or interfaces, fostering unique discoveries~\cite{willett2007scented, matejka2013patina}. 
This prior work is relevant for our research, as we seek to design embedded discussions that can better facilitate engagement with data visualizations and across different users.
\section{Anchored Comments and Discursive Patinas}
\label{sec:design}
This research investigates how comments on data visualizations could facilitate a critical and constructive discourse about a subject, its visual representation, and the underlying data. Based on the assumption that anchoring comments in the visualization could facilitate a well-grounded discourse, we seek to close the gap between visualization and discussion, which translates to three design goals:

\noindent\textbf{DG1---Anchor comments in visualization.}
Comments should be directly embedded as annotations in a data visualization, specifically targeting particular regions. The annotation mechanism should be type-agnostic to allow any kind of visualization to be commentable. A comment's visual appearance should convey its scope, i.e., location and spatial extent in the visualization. To cater to a wide range of users with varying visual literacy, from casual viewers to professional analysts, the commenting interface needs to have a low barrier to entry, making it straightforward to both read and add comments.

\noindent\textbf{DG2---Provide discursive patina.}
The interface should accommodate comments at scale by effectively managing, displaying, and aggregating a large volume of comments in a way that gives an overview of the discourse unfolding on the visualization. The discursive patina emerging from the anchored comments should indicate where interactions between viewers are most concentrated, for example, guiding viewers to areas of high engagement or controversy within a given chart.

\noindent\textbf{DG3---Balance discussion and visualization.}
While offering a visual synthesis of all comments, the superimposed visualization of the discussion should preserve legibility of the underlying visualization. The challenge is to integrate comments in a way that enhances, rather than detracts from the visualization. A balanced approach needs to ensure the primary function of the visualization, while enriching it with a discursive layer.

We address DG1 by allowing users to draw semi-transparent rectangles---anchors---onto the visualization to designate the region of the visualization their comment relates to.
We address DG2 by styling those anchors according to data about their comments: number of replies and likes, type, timestamp, resulting in meta-visualizations, each showing how different regions of the visualization are referenced in the comments.
Finally, we address DG3 by providing interactive control over the visual presence of the discussion and the particular characteristics that are being emphasized (e.g., category, popularity, or relations).
\platform{} is implemented as an open web-platform supporting the ability to upload visualization images, leave and browse comments, reply to others' comments, and explore discussions through different discursive patinas.

While designing these solutions, we addressed a range of problems related to \textit{Which information to collect and display about the discussion? 
How to collect this information from user comments?
How to visualize all the information in a clear way? \emph{and} 
How to best balance visualization of information about the discussion with the legibility of the underlying visualization?} 
In the remainder of this section, we explain the design, rationale, and solutions we found for anchored comments (Sec.~\ref{sec:anchoredcomments}) and discursive patinas (Sec.~\ref{subsection:patinas}).

\begin{figure*}[h!t]
 \centering
  \includegraphics[width=1\textwidth]{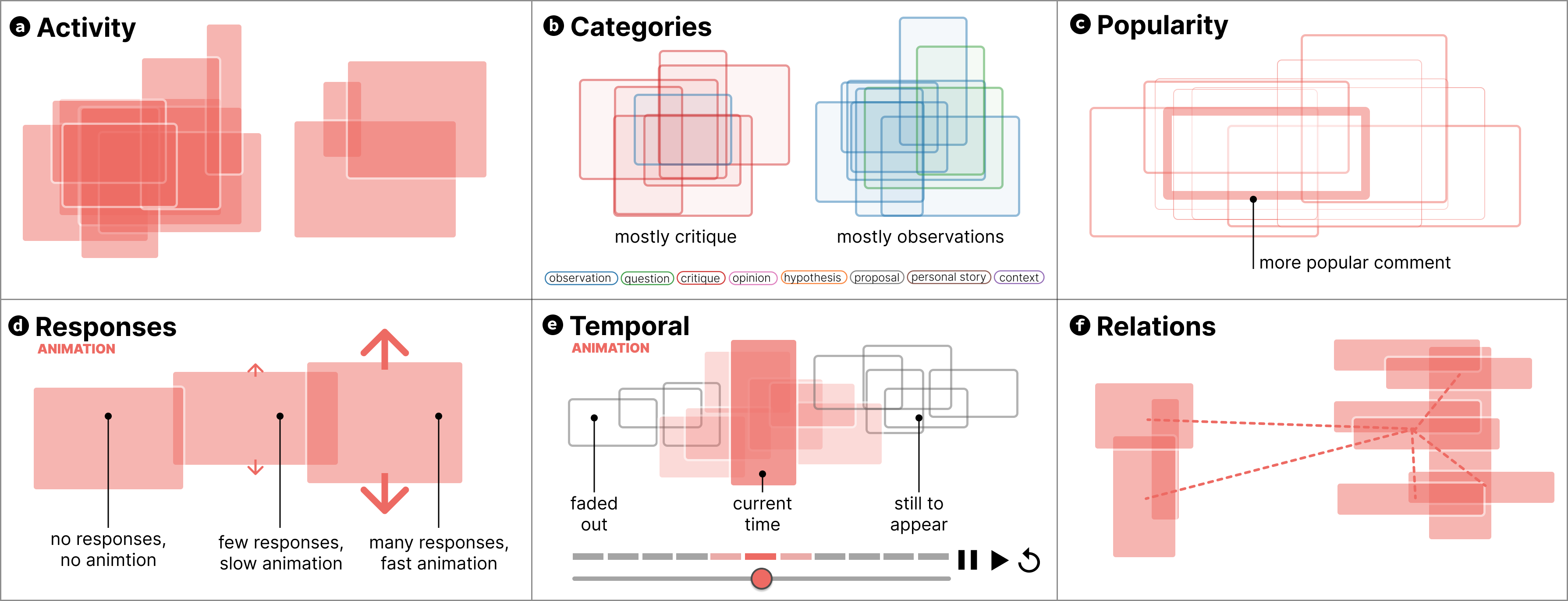}
  \caption{\label{fig:patina}Illustrations of the six patina encodings: each encoding is driven by different metadata from single anchored comments.}
\end{figure*}

\subsection{Anchored comments}
\label{sec:anchoredcomments}

\noindent\textbf{Drawing anchors.} Comments are created by dragging the mouse over the region of interest in the visualization to create a rectangle selection (Fig.~\ref{fig:authoring}a). Pixel locations and dimensions of these anchors are recorded in a database, coupling the comment with the visualization (similar to PixelClipper~\cite{walny2020pixelclipper}). Each comment in \platform{} must have at least one anchor. Additional anchors can be drawn and associated with a comment in order to relate two or more regions of interest to that same comment (see Sec.~\ref{subsection:patinas}).

\noindent\textbf{Writing comments.} Upon lifting their finger after drawing an anchor, users are presented with a form (Fig.~\ref{fig:authoring}b) to enter their comment. A thumbnail refers to the regions they just selected in the visualization. The form \revised{records the following information:
\textit{a)} user name (optional); 
\textit{b)} the user's comment (optional, no length limit); and 
\textit{c)} a group of optional categories to classify the comment.}
Categories for comments are inspired by a recent study classifying types of comments Reddit users posted in relation to visualizations~\cite{kauer2021public} and include: \textit{observations, hypotheses, questions, critique, context, personal stories, opinions}, and \textit{proposals}. These categories aim to capture structured data about the meaning of a comment. Immediately after pressing the `submit' button, comments are published on the web platform open for anyone to read them (Fig.~\ref{fig:authoring}c).

\noindent\textbf{Browsing and reading comments.} 
Comments are displayed as part of elements that include: the full comment text, the author's name, the selected category, the number of likes and responses, and the creation time. When hovering an anchor, the associated comment appears next to the anchor. All comments appear in a scrollable list on the right of the visualization (Fig.~\ref{fig:authoring}), with an additional thumbnail that previews the anchor. Hovering over a comment on this list highlights the associated anchor on the visualization. Clicking on a comment in the list or on an anchor on the visualization shows a conversation view of the comment on the right side, which includes the original comment and a threaded view of all its replies.
The list of all comments can be sorted by date, popularity, and number of responses. By default, comments are sorted chronologically, with the newest comments on top. Switching to another patina encoding shows popularity scores or number of responses per comment sorts comments by those measures.

\noindent\textbf{Responding to comments.} 
The conversation view features a form to reply to comments, which includes
a) the respondant's name (optional) and
b) a text comment (optional, no character limit). Replies are explicitly linked to a
comment and hence implicitly linked to the anchor of the original comment. All replies refer to the original comment; replies to replies are a future extension. Another way for a reader to express that they liked a comment is to press a heart-shaped button next to the comment, increasing its popularity score.

For each comment (excluding replies), the following meta-data are collected:
\textit{a)}~anchors and their pixel coordinates in the visualization,
\textit{b)}~categories selected by the user, 
\textit{c)}~number of likes,
\textit{d)}~number of replies, and
\textit{e)}~date and time the comment was created. These aspects are the basis for the design of different patina encodings.

\subsection{Discursive patinas}\label{subsection:patinas}
All comment anchors are visualized as semi-transparent rectangles with a minimum opacity of 5\%. This allows anchors to overlay each other, showing where comments accumulate and which parts of the visualization are most discussed. All anchors are in turn grouped and the group opacity is set to 50\%. This ensures that the underlying visualization will remain legible in the presence of large numbers of overlapping anchors and therefore remain readable (DG3). To further support the legibility of the underlying visualization, we lower the saturation of the chart to 30\%. This enables a balance between allowing strong colors of the visualization to remain visible (e.g., to identify categories or color scales) while also providing visual prominence to the colors of the anchors.
To visually identify comments on the visualization, each individual anchor has a stroke to signal the size and position. To help select anchors among many, smaller anchors are placed on top of larger ones.

Data about comments are encoded through visual variables of the anchors. To avoid too much information being encoded in the same visual mark, we designed six different encodings, leading to six different types of patinas. We mapped one facet of comment meta-data to one visual attribute per patina. We chose different visual attributes to give each patina an individual identity.\footnote{The supplemental material includes comprehensive examples of patina encodings.}

The default \textbf{Activity} patina encodes the parts of the visualization comments refer to. Comment anchors are colored red with an opacity ranging between 50\% and 5\% according to the number of anchors on the visualization. Overlapping patinas show which areas of the visualization are commented on most, i.e., spark most discussion. Fig.~\ref{fig:patina}a illustrates how the area on the left attracts more comments than other areas, leading to a more salient color in that area.

The \textbf{Category} patina encodes the eight types of comments that users can use to tag their comment (\textit{observations, questions}, etc.). This allows readers to gauge what people are saying about different regions in the visualization. Each category is mapped to the color of the anchor. Given the different hues, we further reduced transparency of the anchor fills. Fig.~\ref{fig:patina}b illustrates how clusters of similar categories are shaded in the assigned colors.
The view can be filtered to only display single categories, e.g., to only view comments tagged as ``critique'' and so identify which parts of a visualization are often criticized.

We measure \textbf{Popularity} of a comment with the total number of likes.
This measure is mapped to the anchored comment's stroke-width (Fig.~\ref{fig:patina}c) ranging from 1px (least popular) to 10px (most popular). Anchors in this view have a white fill (5\%) to put more emphasis on the anchor outline. This patina shows which parts of the visualization and discussion result in the most positive reactions. For example, this patina could show if people agree or like a particular pattern, feature, or information in a visualization.

The \textbf{Responses} patina visualizes the distribution of replies as a measure of quantity of the discussion happening and shows which parts of the visualization are most discussed, potentially showing controversy (`hot areas'). The number of replies per comment is encoded as the amount of animated jitter of an anchor (Fig.~\ref{fig:patina}d): more jitter and movement indicates more replies. Otherwise, anchors use the default encoding (5\% red fill; 50\% red outline). 

The \textbf{Temporal} patina represents the dynamics of discussions. As more anchors pile up onto the visualization, it would be possible to lose information about the unfolding and different phases of a discussion.
The temporal patina is an animation that fades anchors in and out in the order of their creation (Fig.~\ref{fig:patina}e). When displayed, anchors use the default encoding (5\% red fill; 50\% red outline).
All comments are grouped into ten segments based on their creation time. The animation cycles through the segments and filters the anchors present in the patina encoding. Comments are faded in during their corresponding cycle, remain for one cycle, and are faded out after that. An additional interface provides a progress bar and three buttons (start, pause, replay) to control the animation.

The \textbf{Relations} patina indicates the connections among multi-anchor comments. Some comments on a visualization may involve more than one region of interest, e.g., when commenting on the application of a color value from a legend to a visual element in the visualization, or when comparing two visual patterns (Fig.~\ref{fig:patina}f). 
The relations patina shows anchors belonging to the same comment as visually linked by a dotted red line. The resulting patina can reveal interdependencies spanning a visualization.

Lastly, users can also select to view\textbf{ no patina} at all, either by selecting the 'none' option from the encoding list or by disabling the 'show comments' toggle above the visualization (Fig.~\ref{fig:authoring}). As a result, no anchors are shown and the visualization is displayed with full saturation. 
\newcommand{\numworkshops}{10}

\newcommand{\education}{[education]}
\newcommand{\group}{[group]}
\newcommand{\comparison}{[comparison]}
\newcommand{\supp}{supp. material}

\section{Evaluation}
\label{sec:evaluation}
In evaluating anchored comments and discursive patina, we want to answer the following research questions:\\
\noindent Q1: \textit{How are anchored comments used and what do people annotate?} \\
\noindent Q2: \textit{How do discursive patinas shape the experience of the discussion?} \\
\noindent Q3: \textit{What are possible applications of discursive patinas?}

\revised{We designed three complementary studies with 10 sessions and a total of 90 participants, each providing a different angle on our research questions. The complementariness was necessary to account for the fact that visualization annotation is very open process, that can require people to get familiar with the topic, may in fact require some expertise, and aims to reflect diversity in the comments and topics.
Moreover, the amount to which people engage with visualizations and actually leave comments can vary strongly among participants,
their relation to the topic,
and general ability to express themselves or participate in discussions.
Consequently, the three studies differed in their
objective (understand open annotations and motivations, understand collaboration, comparison with baseline), and have different setups informed by:
the context of the discussions (education, shared discussion, guided),
participants (domain experts, students), 
task (open annotation, collaboration), and 
scaffolding (controlled, open).}

\subsection{Open annotation sessions \revised{\texttt{(OA)}}}
\revised{The first study aimed to understand our interface in an open annotation and education setting.}
We invited ten participants (\texttt{OA1} to \texttt{OA10}) to a 60 minute session and gave them tasks and exploration prompts. Unique to this format, participants \textit{were not} familiar with the charts presented to them. We selected three publicly available visualizations portraying topics covering age distributions, recipe ingredients and economic data (see \supp{}).

Additionally, we invited 60 undergraduate university students (\texttt{OA11} to \texttt{OA70}) attending an introduction to data science and visualization course to use \platform{}. Over four consecutive weeks, we uploaded four different publicly available visualizations (see \supp{}) to the platform and presented them to students. We introduced the platform features and asked students to use \platform{} to comment on the week's visualization. During class time, we encouraged them to explore anchored comments and discursive patina, and asked them to conduct a close-reading of the visualization: exploring the data, generating insights, posting questions, and following their peers' comments. We also invited students to asynchronously use \platform{} throughout the week, for those who wanted to contribute after the allotted class time had ended.
After concluding all four sessions, we analyzed the interaction logs and the participants' comments.

\subsection{Domain expert workshops \revised{\texttt{(DE)}}}
\revised{After the first study, we felt that we lacked understanding of how \platform{} could support more nuanced discussions about visualizations. So for our second study, }we invited domain experts from two university research groups to two separate workshops (total 13 participants) (\texttt{DE1} to \texttt{DE13}). 

The workshops were conducted online via video calls, lasting one hour. \revised{We used these workshops to understand} how experts used anchored comments (Q1) and read the discursive patina (Q2) when they engaged with visualizations from their own research and domain.
In each workshop, we presented three data visualizations which either had been created by participants, used in their work, or reflected the subject of their research (see \supp{}). One group (10 participants, 1 moderator) consisted of experts from law, design, and peace research who discussed charts and dashboards about peace data. The other group (3 participants, 2 moderators) consisted of experts from art history, art, and design and  discussed charts from cultural collections and data journalism. We briefly introduced participants to the platform's features, including drawing anchors, authoring comments, and switching patina encodings. Workshops were structured with two tasks that directed the attention of participants to the anchored comments and discursive patina functionality. Specifically, participants were asked to: explore the visualizations and annotate using anchored annotations; and to interact with and understand the patina. 
 
For each task, participants had 20 minutes to use \platform{}. The workshops ended with a 20 minute group discussion moderated by the first author, asking for feedback and reports of participants' experiences. In the discussion, we asked participants where else they could imagine using anchored comments and discursive patinas (Q3). After concluding both workshops, we analyzed the transcripts using an inductive approach to identify patterns in participants' statements.

\subsection{Comparative evaluation \revised{\texttt{(CE)}}}
\label{sec:compare}
The final study aimed at understanding both the impact of anchored comments on participants' reading experiences (Q1) and how the discursive patina influenced participants' engagement with the underlying visualization and associated discussions (Q2). In this \revised{controlled} format, we held three sessions with a total of 11 participants.

\begin{figure}[h!]
  \centering
  \includegraphics[width=\linewidth]{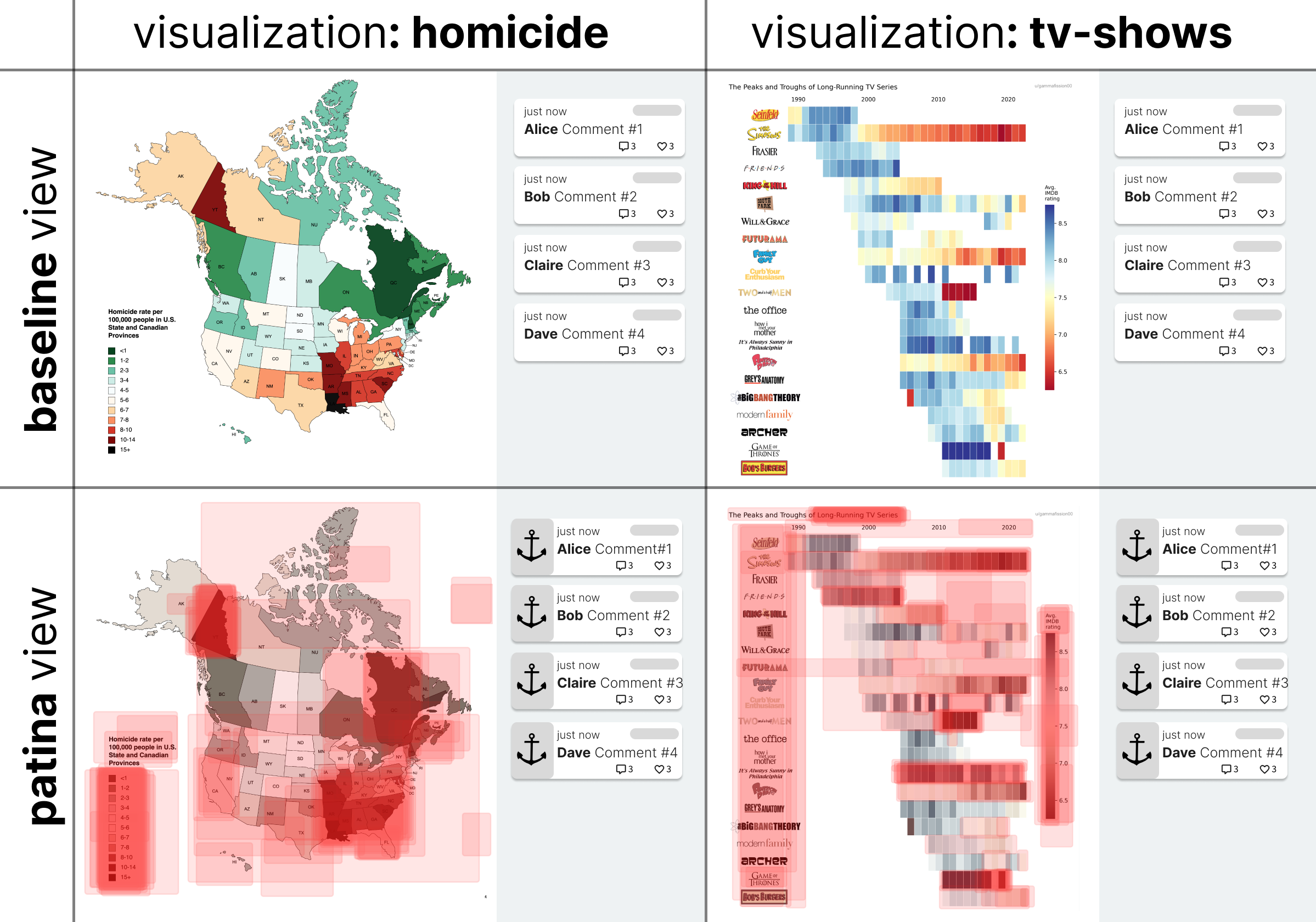}
  \caption{The different viewing conditions for the comparative study show the same two visualizations and associated discussions with varying levels of visual integration. \textit{Top Row} --- Baseline view: Comments are alongside the visualization with no interaction link to the visualization. \textit{Bottom Row} --- Patina view: Comments are anchored onto the visualization and a discursive patina emerges from overlapping anchors. }\label{fig:reddit-visualizations}
\end{figure}

\revised{We compared \platform{} to a baseline interface designed to mimic common interfaces for visualization discussion and annotations (e.g., on social media, news outlets, or research projects~\cite{viegas2007manyeyes, walny2020pixelclipper}). The baseline interface \textit{only} displayed comments juxtaposed to the visualization. Comments were ordered temporally, but could be sorted by popularity, number of responses, and filtered by categories}. 
To achieve this,
we used our \platform{} implementation and \textit{removed} all functionality for anchoring, patinas, and thumbnails from comments (Fig.~\ref{fig:reddit-visualizations}).

For the visualization examples, we turned to the /r/dataisbeautiful community on the social media platform Reddit to find visualizations that were discussed extensively. Without a large volume of discussion, the patinas are not visually obvious. Using pre-existing visualizations with comments balanced the necessity for many comments that were also realistic in origin. Also, comments from this source remain meaningful in the baseline condition, as their comment text usually includes a reference to a part of the visualisation. The first author, who frequents and contributes to the community, chose an initial set of potential visualizations. Along with the rest of the authors, we narrowed our decision down to two visualizations that had discussions pertaining to different parts of the visualization.

The two selected data visualizations were homicide rates in North America 
and popularity ratings of TV shows 
(see \supp{}).
Each visualization had more than 250 comments and responses, which were scraped with the Reddit API.\footnote{\href{https://www.reddit.com/dev/api/}{Reddit API}} 
The first author went through each comment, determined which area of the visualization the comment referenced, and created a corresponding anchored comment in \platform{}. For example, comments that mentioned the color scale were mapped to the color scale legend in the visualization; comments that mentioned specific data items, like states or TV shows, were mapped to contain the respective marks in the visualization. In twelve instances, comments could not be attributed clearly to a specific area in the visualization and were excluded from the study. For each comment, we also transferred the original upvote scores, the number of comment responses, and the creation timestamps from Reddit onto \platform{}. Each comment was further tagged with one of the established categories. To ensure similar numbers of annotations on both visualizations, we stopped after the first 100 comments, with the combined number of comments and responses being 250 for each visualization.

We recruited 11 participants (\texttt{CE1} to \texttt{CE11}) through an open call promoted on social media and mailing lists across the authors' respective institutions. We conducted three sessions online via video with varying participant numbers (1; 5; 5 participants per session). Sessions did not require collaboration among the participants and were only carried out with multiple participants for efficiency.
Each session was run by the first author, lasting 55 minutes on average. The participants first saw one of the two visualizations in the baseline view. They were prompted to explore the visualization and comment what stood out to them. 
Once participants completed this task, they were shown the second visualization in the patina view with the same instructions. There was no time limit for the tasks and we asked participants to report once they were done. Across sessions, the order of visualizations was alternated to account for any learning bias. 

After the tasks were completed, participants were directed to a structured online survey with questions about navigation strategies, notable observations, and reactions to anchored comments and the discursive patina (see \supp{}). We concluded each session with a group discussion that lasted 20 minutes. To address Q3, we used the discussion to ask participants how they would use the technique in other contexts. After concluding both workshops, we analyzed the survey responses and participants' statements.

\newcommand{\numparticipants}{90}
\newcommand{\numentries}{263} 
\newcommand{\numcomments}{225} 
\newcommand{\numresponses}{38} 
\newcommand{\numbaseline}{21} 
\newcommand{\numerror}{8} 
\newcommand{\numanchored}{196} 
\newcommand{\numanchors}{212} 
\newcommand{\result}[1]{\texttt{R#1}}

\section{Results}\label{sec:results}
\revised{Across all studies, we collected comments from participants through anchored comments (\includegraphics[height=1.3\fontcharht\font`\B]{figs/comment.png}), 
verbal recordings from think aloud and discussions (\statementsymbol), 
and the post-hoc questionnaire (\includegraphics[height=1.3\fontcharht\font`\B]{figs/survey.png}). The open annotation session (\texttt{OA}) yielded 77 anchored comments (65 of which in the timed session and 12 from the classroom setting). The domain expert workshops (\texttt{DE}) yielded 78 anchored comments.
The comparative evaluation (\texttt{CE}) yielded 41 anchored comments. 
Across all studies, we collected interaction logs. In the following, we refer to participants by the study abbreviation, the participant number in that study, and the type of comment we cite, e.g., \includegraphics[height=1.3\fontcharht\font`\B]{figs/comment.png}~\texttt{DE6} is an anchored comment created by participant \#6 in the domain expert study. No participant number indicates an anonymous submissions. 
}

\subsection{Writing and reading anchored comments}\label{subsection:resultsq1} 
Working towards answering Q1, we present five results (R1-R5) that show how anchored comments are created and used for visualization interpretation. 

\begin{figure}[h!]
  \centering
  \includegraphics[width=0.8\linewidth]{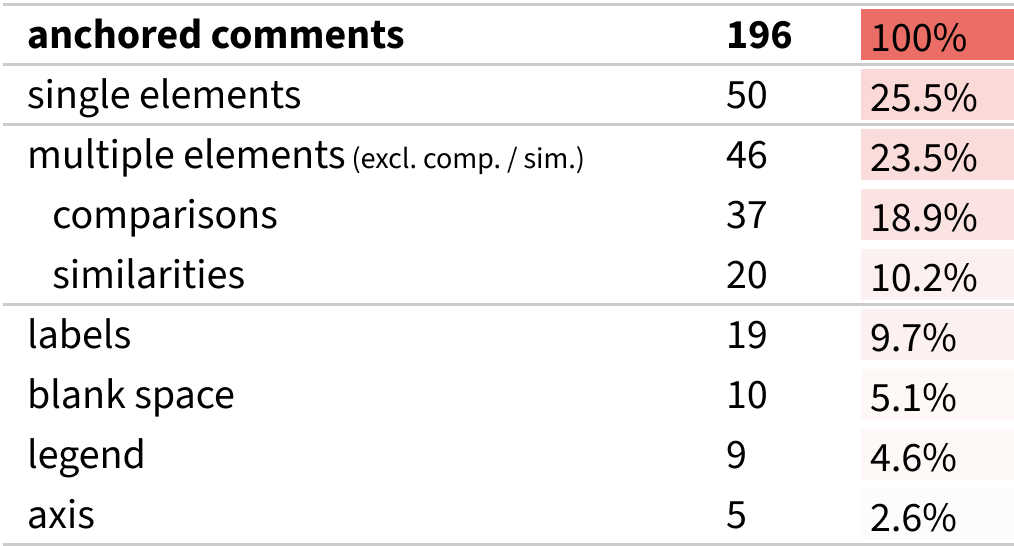}
  \caption{Open coding of anchor usage, analysing what areas of the visualization are marked with anchors}\label{fig:anchor-usage}
\end{figure}

\noindent\textbf{R1. Anchors highlight details, insights, and context.}
The vast majority (86\%) of anchored comments use one anchor, the remaining comments use two or more anchors, resulting in a total of \numanchors{} anchors. We qualitatively analyze anchored comments in an open coding process. In three coding passes by two authors, we read each comment and its associated anchors to come up with and converge on eight categories that describe which parts of a visualization the anchors relate to (Fig.~\ref{fig:anchor-usage}).

Most comments relate to single (26\%) or multiple visual marks (52\%) in the charts, allowing participants to comment on the data or its representation. We observe how some comments that refer to multiple marks express insights about similarities and differences: Comparisons highlight changing trends in the data, e.g.\comment{Interesting to see how Washington and Oregon are different}{CE}; similarities point out elements that look alike, e.g.\comment{Estonia, Latvia, Lithuania all look so similar}{OA1}. In both cases, single anchors are used to refer to elements that are next to each other while multiple anchors link elements that are further apart on the visualization.

While rare, comments on blank space (5\%) allowed participants to express interesting opinions that were otherwise difficult to associate with a single part of the visualization. The comments in these areas mainly pertained to three topics:
questioning the absence of elements, e.g.\comment{I wonder why this is empty. No data? Design choice?}{DE};
proposing missing elements that they wanted to see in the visualization, e.g.\comment{place for a legend?}{DE8};
or placing comments that are unrelated to the anchored area just to place them somewhere. 

Anchors also pointed to contextual elements surrounding the visualization: Participants commented on labels (10\%) e.g.\comment{I find this wording a bit cryptic}{DE4}; legends  e.g.\comment{definitely too many colors and hard to see. why not just add the number??}{CE7}; and axes (3\%) e.g.\comment{Does the horizontal axis indicate passing of time?}{DE7}.

Analyzing anchors' sizes, we learn that anchors are generally used to refer to relatively detailed parts of the visualization with the the median anchor size spanning 12\% of the width and 11\% of the height of the visualization. 

\noindent\textbf{R2. Comments guide visualization interpretation.} 
Comments helped participants make sense of the visualization in three ways: by reducing the barrier to entry to understand the data and visualization, adding relevant context, and introducing alternative interpretations.
Comments captured other people's preceding efforts to understand and explain the data, which other readers then could use and consequentially comprehend without closer examination of the visualization: \statement{I looked at the most popular comments and found this experience quite compelling as a way to find interesting factoids without having to `put in the work' of looking through the vis myself}{CE3}.

Comments also guided the readers' attention to novel or hidden information about the visualization design: \statement{Sometimes people don’t see things that are there already - like a hidden legend – or design decisions like the blue and the red it’s not fully explained}{DE10}. Together, comments about the data and visualization acted as scaffolds to new readers.
Comments added context that otherwise would remain unknown: 
\statement{[Comments] focused me on others' observations and surfaced things I wouldn't have learned otherwise}{CE3}. 

Across comments, some were pertinent to the data, while others were more personal. One participant reflected on how they used the personal stories to make sense of the visualization: 
\statement{All the personal background stories [...] gave me some information about hypothesis why certain things were colored a certain way. In the end, the graph itself doesn't answer those kinds of questions, since it only showed the numbers}{CE7}. Similarly, reading other people's interpretations of the data sparked reflection on assumptions participants had about the data: \statement{It’s a nice opportunity to challenge what you think you know about a process or how you think you understand it. [...] It’s nice to challenge your own assumptions}{DE7}.
Considering that comments don't require participants to \statement{put in the work}{CE3}, an important caveat is the risk of bias. One participant acutely related their lack of effort, foregoing `put[ting] in the work', at the expense of taking others' interpretation as accurate: \statement{The page is already showing me the impressions of other people. It does not allow me to first form my understanding and it’s visually biasing me}{CE6}.
Ultimately, whether the comment supported an easier or more nuanced reading of the visualization or introduced productive discourse that sparked reflection, comments affected how the visualization was interpreted by new readers. 

\noindent\textbf{R3. \textit{Observations} and \textit{questions} are prominent comment types.} 
For 71\% of comments, participants selected a category; the most used categories were \textit{observations} (27\%), \textit{questions} (15\%), and \textit{critique} (10\%). Others were opinions (6\%), hypotheses (5\%), proposals (5\%), personal stories (2\%), and background information (2\%). 

Observations ranged from simple descriptions of visual marks, e.g.\comment{Crazy shape}{OA1} to conclusive insights into data similar to aforementioned comparisons, e.g.\comment{most activity in this time frame}{DE12}. Questions covered a broad range of topics from asking for clarification of data, e.g.\comment{which demographic of people were used for this?}{OA}, reasons behind visual patterns, e.g.\comment{Do the blank spaces tell us something?}{DE10} or \comment{why is this shift here?}{DE}; or explanations of the encoding, e.g.\comment{are the colors explained somewhere?}{DE8}. Critical comments primarily point out frictions participants had when trying to make sense of the visualization, e.g.\comment{I don't get what this dotted line is about}{OA}, but also raised awareness of questionable data or encodings, e.g.\comment{I think this is highly biased because it's just based on this one source and not a more broad survey}{OA}.  

Participants used observations to point out \revised{areas of interest, }e.g.\comment{The first thing that catches the eye is that the markers are populating a certain time frame [...]}{DE10}, but also to elicit more information from others, e.g.\statement{I posted some observations but did not really know their context so Its useful to see what [domain expert] makes of them}{DE12}. Others asked questions to interrogate the visualization and guide their own sensemaking process, e.g.\statement{My comments [...] were mostly questions because i didn’t fully understand what [the visualization] is supposed to be telling me}{DE7}.
Most responses are reactions to the two most frequent categories (40\% respond to questions, 30\% respond to observations), with a disproportionately high engagement on questions that make up 15\% of comments but attract 40\% of responses. 

\noindent\textbf{R4. Anchoring changes how comments are written.}
The structured survey shows that a majority of participants agree (46\%) or strongly agree (27\%) when asked, whether anchoring comments changed their commenting behavior. 
Participants reported that the presence of the patina reduced the amount of text that they wrote in their comments. We confirm this by comparing text lengths across comments in the separate conditions. The median comment length was 36\% shorter when comments were written in the patina view (91 characters) compared to the baseline view (142 characters). In some instances, the reason was practical: the participant needed less text to describe what part of the visualization they were commenting on. Participants found the coupling between the comment and visualization useful because it prevented the need to \statement{write an essay}{CE2} and instead made it \statement{possible to comment, e.g., a specific row without describing it in words}{CE2}. 

Additionally, the act of choosing where to anchor a comment changed the types of comments participants considered appropriate to write. Participants correlated the position of the anchor with spatial significance, which made it difficult when they wanted to leave a general comment: \statement{Since I had to select specific areas, I did not really feel like I could make general comments even if I wanted to. I had to think about comments that would make sense for a specific area of the visualization (or that was pertaining to a specific area), so I kind of restrained myself from making some comments that I think I would have made otherwise}{CE9}. In this instance, the generality of the comment clashed with the specificity of the anchor, which prevented the participant from leaving the comment.

\noindent\textbf{R5. People's engagement varies across visualizations and settings.}
Within \platform{}, there was a variety of visualization types for participants to discuss. We observed that some visualizations attracted more comments than others. Among the visualizations that elicited more comments were visualizations with small multiples, dashboards, and maps. We describe these visualizations as \textit{information dense} as they describe a large number of entities, carry a lot of information, and can invite close inspection. Participants reported that information density fostered their engagement: \statement{Having lots of data points is more overwhelming but maybe it’s easier to annotate because there’s more things to take note of}{DE12}.
\revised{Further, the setting of individual user studies had an impact on engagement: For example, the open annotation session \texttt{(OA)} with students yielded 0.2 anchored comments per participant, while each domain expert \texttt{(DE)} on average wrote 6 anchored comments.}

\subsection{Discursive patinas for navigating discussion}\label{subsection:resultsq2} 
Answering Q2, we observed two recurring themes in how participants used the discursive patina to explore and navigate the discussion. While these themes were mentioned across all evaluation studies, we primarily based our results on the comparative evaluation which best highlighted the effect of the patina view compared to the baseline view.

\noindent\textbf{R6. Patinas help navigate discourse.} 
Participants described the discursive patina as a
\survey{map},
\survey{heatmap}, and
\survey{meta-visualization}
that helped gain an overview of the discussion: \statement{some areas were `more red' than others, indicating that there were more comments made on that specific area of the visualization}{CE9}. With the overlapping anchors creating a salient visualization of comments, participants used the saturation of color and overlapping areas to find where comments were concentrated. One participant remarks on how the patina illustrates the overall perception of the visualization, stating: the patina is the \survey{reception of the visualization among the readers}
that answers, \survey{what are the hottest findings}.

The baseline view, in contrast,  made gaining an overview of the discussion difficult. One participant noted the chaotic nature of connecting discussions to the visualization:
\survey{I don't think I was able to analyze the discussion itself, for it looked to me like a rather disorganized discussion forum}.
Other participants also found the baseline view particularly taxing. They described difficulty in locating what parts of the visualization the comments referred to. This was articulated by one participant who discussed the challenges of both trying to understand the visualization and also the comments concurrently:
\survey{in the [baseline] interface it was really hard to pinpoint the particular discussion points. you are losing the visualization out of your sight and instead focus on the comment section}.

While all participants in the comparative sessions were prompted to explore different patina encodings (Fig.~\ref{fig:patina}) only one mentioned them in the survey: \statement{to see what is most commented on, what is liked the most, and how things are linked. This is something completely new}{CE6}. Based on the interaction logs, we learn that all comments have been written in the default `activity' encoding and we do not know whether participants used different patina encodings to more successfully navigate discussions.

\noindent\textbf{R7. Patinas can overwhelm and obstruct.}
Although the discursive patina helped users navigate discussions, our analysis also shows that the patina impeded their interaction with the underlying visualization. One participant wrote a comment using the interface to point this out: \comment{I'm reading comments and then I want to see the data but it's a bit hidden by the overlay so it's a lot of back and forth between the visualization with and without comments}{CE}. Others reported being overwhelmed by the patina: \survey{when you are entering the page the first thing you see are already the red annotations}.
Two participants decided to ignore the patina because they \statement{found that frustrating [and] switched to the list}{CE3}, or switched off the [patina] view altogether in order to \survey{see the vis properly}.
They reported concerns that the patina may have distracted them from fully looking at the visualization: \survey{I didn't actually look through the viz, so maybe there are things I might have noticed or cared about that I didn't see?}.
In addition to visual obfuscation, participants also mentioned that they felt biased by the content of the patina. One participant reflected on how the patina prevented them from developing their own interpretation of the visualization: \survey{because the page is already showing me the impressions of other people, it does not allow me to first form my understanding and it’s visually biasing me}.

\subsection{Scenarios}\label{subsection:resultsq3} 
Finally, we asked participants to speculate about scenarios in which they would want to use the \platform{} interface (Q3). To answer this question, we report participants' statements and group them into three scenarios that arose during the discussions.

\noindent\textbf{Scenario 1: Visualization design feedback.} Participants suggested using the platform to provide feedback on visualization design through a productive dialogue between visualization designers and other team members.
On the one hand, this would allow the team to point out friction points in the design and iterate, e.g., a visualization's encoding, layout, or labels. On the other hand, visualization designers could use the feedback to better understand the underlying data: \statement{This could be used to talk about mock-ups, understand underlying data and get comments from your peers before you publish something}{DE11}. 
One participant specifically talked about anchored comments as a precise tool to pinpoint design critique: \statement{In news outlets, you have very little comments about design critique. This interface is sparking this discussion, because comments are so localized}{CE6}. 
The idea of design feedback was particularly popular in the domain expert workshops, as the visualizations in this format came from the participants and echoed prior design feedback processes that they had done together. 
One participant mentioned, how the platform could streamline this process and help with facilitate a dialogue with relevant stakeholders: \statement{A really useful application for this would be a dialogue with our funders. This would be a useful, quick, and interactive way for them to ask questions and discuss future development instead of writing big lists of questions to go through in meetings}{DE6}.
This scenario came up in a workshop, when the discussion helped a group's designer understand overlooked domain knowledge and its relevance to the visualization: \statement{Because I’m part of the creative team for this visualization it’s really interesting to see this feedback because there is domain knowledge i don't know and it’s really relevant for my work}{DE9}.

\noindent\textbf{Scenario 2: Help with public sensemaking.} As reported in R3, we found, many instances in which participants used the interface to pose questions and get answers. R2 shows how this dialogue can support readers in their sensemaking. 
Participants envisioned \platform{} as a public forum to discuss data visualizations. In this scenario, one participant mentioned how \statement{one answer allows you to see the visualization in a whole other way}{DE10}.
This public dialogue can be particularly powerful if data experts are part of it. One participant mentioned how they used the platform to elicit context from domain experts to better understand the underlying data: \statement{I posted some observations but did not really know their context so it’s useful to see what [Domain expert] makes of them}{DE12}.

\noindent\textbf{Scenario 3: Local knowledge collection.}
Participants envisioned how discursive patinas could support collecting personal stories from a variety of sources on the ground. Participants deliberated both on the opportunities and drawbacks of such a use case. While they saw the potential for anchored annotations to facilitate discussion amongst people with different perspectives, they also acknowledged that discussion moderation would become more important. One peace researcher: \statement{If we are talking about limitations, bringing together different experts brings together completely different approaches [and truths]. If we were doing something where we are engaging with different people to [...] feed in their perspectives, you need to carefully think about representation and balance who gets to be a part of that}{DE7}.
The participant noted the opportunity to foster \textit{``ongoing discourse''} with users holding local expertise. They imagined how the visualization could benefit from input provided by people with \textit{``very different expertise from in-country experts''}.
\section{Discussion}

With \platform{}, we set out to design an interface that allows to directly annotate information \textit{on} 
a visualization and thus help relate comments to visualization content (DG1), to help understand an ongoing discussion (DG2), and to balance the presence of the visualization and information about the discussion (DG3). 
The results from our studies suggest that anchored comments are used for detailed comments and to contextualize those comments with information from the visualization~(R1), often in the form of observations, questions, and critique (R3, R4), with detailed and information rich visualizations soliciting the most comments~(R5). Comments, in turn, can guide the sensemaking process and help people reflect on their own opinions (R2). We found that most of our participants valued how patinas helped them gain an overview over the discussion and find entry points to begin exploring the visualization and discussion (R6). Other participants preferred the raw visualization and found patinas competing with the visualization~(R7). 
Participants described scenarios that could benefit from anchored comments and discursive patinas, all of which imply an exchange between strangers and require the ability to read and write comments as well as to form discursive patinas pointing out issues, localizing questions, and contextualizing explanations.

In the remainder of this section, we reflect on these findings and our design goals, \revised{highlight limitations of our approach,} and discuss implications for studying and designing discussion interfaces. 

\subsection{Close and distant readings} 
One immediate observation is that anchored comments support \textbf{close reading of visualizations} as people were more likely to annotate detailed visual patterns and visual marks.
In the humanities, close reading refers to reading individual texts carefully, listening to what the text is saying and its method of delivery. Close reading in visualization is consequently strongly related to visualization literacy~\cite{boy2014principled}, reading visualization techniques, and understanding flaws~\cite{pandey2015deceptive} and visual patterns in visualizations~\cite{wang2020cheat}. However, while visual literacy tests evaluate the understanding of common visualization techniques~\cite{lee2016vlat} and criticality~\cite{ge2023calvi}, those tests usually focus on common charts and question-based assessments. We think that close reading of complex visualization requires a much deeper engagement and open questions such as used in visualization evaluation~\cite{bares2020using}: asking people to openly comment on what they see and observe, what they found most significant, what things meant to people, and what additional information would have been required. Anchored comments can guide people in these activities. 

While anchored comments focus people on the close reading of of visualizations, discursive patinas can be seen as supporting \textbf{distant reading of the resulting discussions}. Again, a method common in humanities research, distant-reading aims to examine large bodies of literature by taking a step back from individual texts and focusing on salient and abstract features across many texts, e.g., accumulation of specific words or phrases, genres, or literary figures~\cite{moretti2013distant}. Similar in our case, discursive patinas can support the distant reading of discussion dynamics by abstracting from textual comments and focusing on data across potentially elaborate discussions. In fact, across the workshops, participants  commented on the discursive patina as a visual guide to the discussion and remarked how it helped them quickly spot where comments accumulated (R2). The different patina types \revised{are desgined to }offer multiple lenses to the discussions to debate questions such as: 
\textit{What are the discussions about? 
Which topics dominate a discussion? 
Are there parts of a visualization that cause more argument? 
Are people actually discussing the visualization? 
How does a discussion unfold over time? }
While there has been considerable research on visualizations for distant reading of texts~\cite{janicke2015on}, discursive patinas are the first attempt to map discussions onto visualization, helping people to zoom in and out of the discussions of a visualization and the information they add.

\subsection{Generalizability, scalability \& expressiveness}
The visual and functional characteristics of anchored comments and discursive patinas as implemented in \platform{} are deliberately independent from the underlying visualization.
We carefully designed the anchors and patinas to ensure some expressiveness in the position and size, while also supporting a wide variety of visualization techniques. 
As noted previously~\cite{lin2022data}, free-form comments may vary depending on the data type (e.g., networks) and thus call for more specific designs to balance expressiveness and scalability. 
With our design, we can point to examples that show up to 100 anchors at a time, which we posit is a scale impossible with simple free-form comments. Future iterations on anchor and patina design could include non-rectangular anchors, general lasso selections, or brushing. We can also imagine annotating individual visual elements or groups thereof, permanently linking comments to those elements.

Likewise, we can think about gathering more data about comments, e.g., assessing stances and topics from the comments texts as common in social media analysis. Such data could inform further user interface components, visualizations, and patinas, showing comments over time, the emergence of topics, or the convergence and divergence of sub-topics within the discourse around visualizations. However, in particular the patinas should be designed in such a way that they either focus on analytical tasks for understanding discussions, or as signposting \textit{some} selected comments and regions as inroads into joining a discussion. Currently, patinas are designed to support both scenarios but future iterations could specialize, taking inspirations from overlaid heatmaps, fog-of-war techniques, and blurs~\cite{kim2017bubbleview} that mask unexplored parts of a visualization, or other traces and autographic designs~\cite{offenhuber2023autographic}.

\revised {\subsection{Limitations}
\textit{Limitations of the patina display: }discursive patinas can \textbf{bias viewers' attention} towards areas with high comment density, overshadowing less-discussed but potentially significant parts of the visualization. This could lead to an unbalanced understanding, where viewers focus more on popular areas rather than interpreting the visualization without any social cues. Likewise, the anchoring mechanism encourages \textbf{local comments} about specific elements of a visualization, rather than high-level comments about the whole visualization.
This could potentially limit the scope of a discussions and prevent broader observations. Eventually, patinas unavoidably \textbf{obstruct the underlying visualization}. However, none of these issues was a design goal of our technique and \platform{} can simply switch off patinas.
}

\revised{
\textit{Methodological limitations:} 
Our complementary studies --- from open and exploratory to more rigorous and controlled --- are limited, as \textbf{not every anchored comment we analyse is created equal}.
Hence, our results are limited by the people participating, the contexts their acted in, and the topic of the visualizations. Still, we are confident that especially the qualitative findings across different contexts and participants provide for a good understanding of the potential of patinas and potential improvements.
In future, we hope to roll out \platform{} over a longer period of time, collaborate with newspapers, and explore the use patinas in more ecologically valid scenarios with more participants.}

\subsection{Critical readings of data visualizations} 
\label{subsec:critical-reading}
One particular design approach to supporting users' navigation of visualizations could be to disrupt the common \textit{seamlessness} of data visualizations. Seamlessness points to the tradition of visualization design that strives towards clean, tidy, and easily digestible data~\cite{dignazio2023datafeminism, kennedy2016work}. As Hengesbach articulates, the problem of seamlessness is that \textit{``it distorts, hides, and disguises the complex reality of data''}~\cite{hengesbach2022undoing}.
She further argues to keep or even add seams to point out i.e., complexities, uncertainty, ambiguity, instead of considering them as potential `tripping hazards' for readers that need to be removed. Seams can provide visual cues that point to the partial and imperfect aspects of data. We saw this attention to seams in R2, when participants reflected on how their interpretations differed from those of others, as comments provided visual cues to these disambiguities.
Anchored comments and discursive patinas can be seen as seams and tripping hazards to visualizations similar to traces in the real-world guiding our use of something: a pre-annotated copy of a hand-book, touch and wear marks on a mechanical device, etc. 

Close reading of visualizations can also include questioning the design decisions behind the visualization and the data collection preceding any visual encoding. In particular, visualizations meant for a broad audience run the risk to make assumptions and statements that might upset people, e.g., missing consideration for accessibility or disputed frontiers on maps. Researchers have already articulated visions for feminist data visualization with its aim to reduce the power differentials between visualization authors and readers\cite{dork2012critical, correll2019ethical, dignazio2023datafeminism}. Such works draw on critical and feminist theories to suggest ethical approaches to data visualizations that expose the materiality of data, reveal the decisions of the designer, and reduce the authority of the visualization. We speculate that discursive patinas could help groups of people collectively unpack the many ways that visualization and the underlying data are historically situated. 
Anchored comments can help pinpoint evidence for such hypotheses.

In that sense, we imagine anchored comments and discursive patinas to play a stimulating role in \textbf{open design processes, }\revised{in which visualization designers benefit from an already established practice to receive qualitative feedback~\cite{Choi2023} from their colleagues.}
While data visualizations often aim for large audiences once published, they are often designed, improved, and discussed by a small, private group of individuals. In contrast, designing in the open invites readers to participate in the creation and iteration of the visualization. We point to successes in open design processes during early moments of the COVID-19 pandemic where visualization authors worked with interested audiences to \textit{``iterate on the original theme, generating thousands of comments, and exceptional levels of engagement''}~\cite{cotgreave2021after}. To facilitate such and other scenarios, we can imagine guided tours~\cite{walny2020pixelclipper} walking readers through visualizations to offer behind-the-scenes insights (e.g., rationale for design decisions), pointing out on-going construction work and asking questions about the reader's comprehension. As potential conversation starters, such designer comments could be visually distinct from other comments.

\subsection{Future directions}
\label{subsec:limiations}
Our results and reflections show that discussing visualizations is far from being an established practice and methodology. Across all studies, it took time for participants to not only read the patinas, but also to feel comfortable expressing their thoughts and engaging with the other comments. To some extent, this might be further illustrated by the fact that participants did not make full use of all patina types (R6). This might also be influenced by the willingness of participants to participate in discussions about visualizations in a study setting. This could be due to a mismatch between the topics and visualizations we chose for two of the studies, compared to the interests of our participants. It could also signal a lack of group cohesion, like those found in online spaces who are designed to attract like-minded people. 

We attempted to rectify both of these constraints by pre-populating \platform{} with extant discussions. To kickstart the discussions---a problem common in these kind of studies~\cite{kauer2021public}---we chose an initial set of comments from Reddit, rather than fabricating comments ourselves. Although Reddit is a source used by others to investigate user engagement~\cite{almahmoud2023vizdat, kauer2021public}, the comments are \textit{not} integrated to the visualization and have hence been created in a different context.

Another promising direction is the support for anchored comments and discursive patina within interactive data visualizations. Interactivity brings interesting technical and aesthetic questions about anchoring comments within specific view parameters as well as the aggregation across a range of display states.

In any case, we identify a major obstacle for visualization research: current visualization design and practices ignore collaboratively reading and discussing visualizations. We see this as an opportunity for visualization research to understand the skills required to conduct \textit{both} close reading of visualization and distant reading of discussions for future critical engagement with data, visualizations, and participation.

\bibliographystyle{abbrv-doi-hyperref}

\bibliography{template}
\end{document}